# Galactic extinction and Abell clusters

R. C. Nichol[1] and A. J. Connolly[2]
[1] *Department of Astronomy and Astrophysics, University of Chicago, 5640 S. Ellis Ave, Chicago, Illinois 60637, USA*
[2] *Department of Physics and Astronomy, Johns Hopkins University, Baltimore, MD 21218, USA*

**ABSTRACT**

In this paper, we present the results of comparing the angular distribution of Abell clusters with Galactic HI measurements. For most subsamples of clusters considered, their positions on the sky appear to be anti-correlated with respect to the distribution of HI column densities. The statistical significance of these observed anti-correlations is a function of both richness and distance class, with the more distant and/or richest systems having the highest significance ($\simeq 3\sigma$). The lower richness, nearby clusters appear to be randomly distributed compared to the observed Galactic HI column density.

## 1 INTRODUCTION

Obscuration due to dust in our own Galaxy is a major systematic problem for most areas of extragalactic astronomy. In particular, it is a severe problem for flux–limited, optical surveys of cosmological objects (e.g. clusters of galaxies) as these surveys are used for statistical studies of the large-scale structure in the universe (Nichol et al. 1992). A spurious clustering signal can be introduced into these surveys via patchy Galactic extinction. This false signal may then be interpreted as evidence for large scale structure.

The Abell catalogue (Abell 1958) remains the most widely used survey of clusters for statistical studies of the cluster large-scale distribution. Abell was well aware of the potential effects of extinction on his catalogue and corrected all his galaxy magnitudes for the estimated effect. He concluded that Galactic obscuration plays a role in the observed distribution of clusters of galaxies and noted that the surface density of clusters decreased rapidly as a function of Galactic latitude, as well as in a few areas of anomalously low cluster density at high latitudes (e.g. an area at $l = 300$ rising as high as $b = +60$).

Subsequent analyses of the Abell catalogue have taken Abell's advice to heart, by including a Galactic latitude selection function ($P(b)$) of the form

$$P(b) = 10^{0.32(1-\text{cosec}|b|)}, \qquad (1)$$

where $b$ is Galactic latitude (see, for example, Abell 1958; Bahcall & Soneira 1983; Postman, Huchra & Geller 1992). Yet few investigators have worried about possible Galactic longitudinal dependences of Galactic extinction (even the anomalous patches highlighted by Abell), and if they have, they have only considered the extreme cases e.g. the large void of clusters reported by Bahcall & Soneira (1982). Furthermore, all investigations of the effects of extinction on the Abell cluster distribution have only implemented internal consistency checks, by utilizing the observed surface density of clusters compared to that expected from a uniform distribution. This does not, however, account for genuine large-scale structure in the cluster distribution which will certainly confuse the issue.

With the recent availability of high quality, large–area independent extinction indicators, like the Stark et al. (1992) HI radio map of the Galaxy, it is now timely to revisit the question of Galactic extinction and its effects on the observed cluster distribution. It is now possible to check for significant anti–correlations between the observed positions and classifications of clusters and the measured obscuration as derived from such independent, external indicators. In this paper, we present the results of such an investigation since the effects of Galactic extinction may represent the largest systematic bias confronting statistical analyses of the cluster distribution and constrains attempts to construct homogeneous, complete samples of clusters.

## 2 HI RADIO DATA

The Stark et al. (1992) HI radio map is the cleanest survey of Galactic atomic neutral hydrogen (HI) presently available. It is complete above a declination of −40 and was constructed using the AT&T Bell Laboratories 20 foot horn reflector at Crawford Hill. The telescope has a FWHM beamwidth of 2° and is relatively free from side-lobe contamination which is a major advantage over previous work. As much as 50% of the measured radiation in a particular direction can be due to contamination from the side-lobes which are typically located tens of degrees away from the pointing direction. Such problems make it very difficult to map accurately the ambient interstellar hydrogen and can lead to false detections (Hartmann 1994). One of the main drawbacks of the Stark et al. map is its' low resolution. However, for the work presented here on the cluster distribution, it is unlikely that this will be a major constraint since the surface density of Abell clusters is low ($\sim 0.1$ per $\deg^2$).

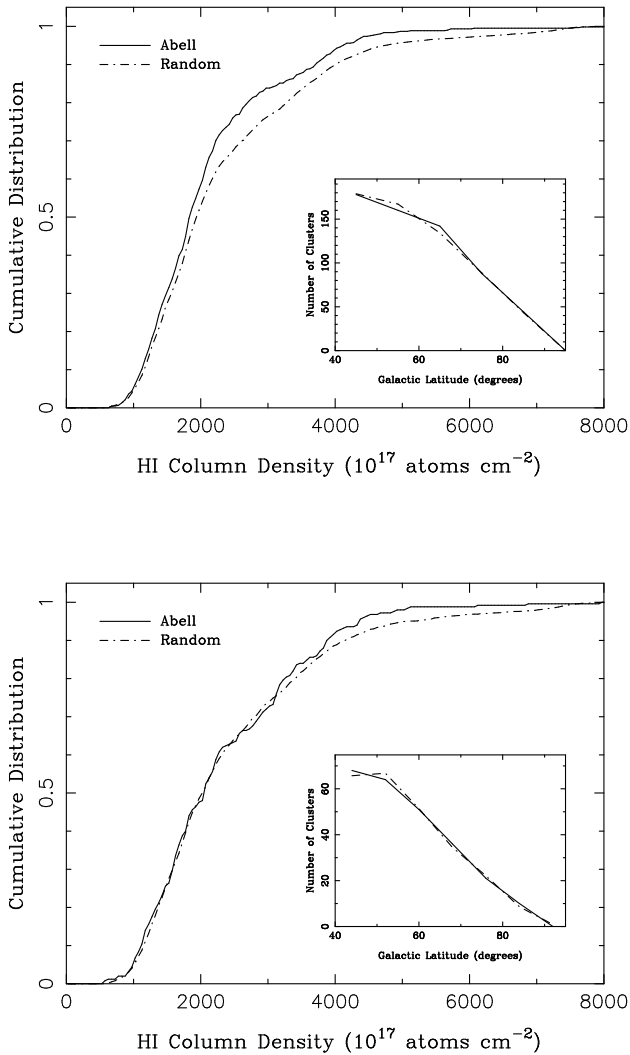

**Figure 1.** The real and random cumulative distributions for the $D > 5$ (top) and $R = 0$ & $D \leq 5$ (bottom) subsamples. It is apparent from these figures that the more distant Abell clusters tend to be at lower HI column densities than that expected from random lines-of-sight, while the lower richness systems appear randomly distributed compared to the Galactic HI. Table 1 contains the results of KS tests on these distributions. The inset plots shows the distributions of Galactic latitudes for these Abell clusters and the random datasets.

The map of Stark et al. was used to interpolate the value of the HI column density (atoms cm$^{-2}$) for any given line-of-sight in the sky north of $\delta = -40°$. A recent comparison of such an interpolation method against data obtained from higher resolution HI observations has shown that it works reasonably well (Elvis, Lockman & Fassnacht 1994). In 80% of the cases, the absolute difference between the interpolated and measured HI column densities is less than $\sim 2 \times 10^{19}$ atoms cm$^{-2}$, or, $\sim 10\%$ of the typical HI values used below. This error is an order of magnitude smaller than the observed spread in HI values seen over the region of the sky considered here and therefore, is expected to have little effect on our results. For the remaining 20% of cases, the distribution of the differences between the two values does show a significant tail with the higher resolution measurements having higher HI values than those interpolated from Stark et al. (the largest discrepancy being $\sim 1 \times 10^{20}$ atoms cm$^{-2}$, or, $\sim 50\%$). These discrepant points appear to have no strong correlation with the observed HI values, although most lie at large HI column densities. Therefore, it would appear that the Stark et al. data, in high HI column density regions, occasionally underestimate the true HI column density. Therefore, if we detect any anti-correlation between the clusters and the HI column densities it will be an underestimate.

## 3 HI-ABELL CORRELATIONS

For the analysis presented here, we concentrate on the original northern Abell survey (Abell 1958) constraining the declinations of the clusters to be above $\delta = -25°$ (the Abell catalogue stops at $\delta = -27°$). Furthermore, we only consider clusters above a Galactic latitude of $b = +40$. This latitude cut is more severe than those suggested by Abell (Table 1 in his paper), since we wish to investigate areas of the sky commonly believed to be free of obscuration effects.

The HI column density values interpolated from the Stark et al. map were used as tracers of the Galactic extinction, assuming a constant dust-to-gas ratio. A typical conversion between HI column density and visual extinction is

$$A_B = 8.65 \times 10^{-22} N(HI) \qquad (2)$$

where $A_B$ is the B magnitude extinction and $N(HI)$ is the HI column density (Nichol & Collins 1994). Only relative extinctions between different parts of the sky are important, since we are not attempting to define the absolute completeness of the Abell cluster catalogue. The validity of using HI column densities, and the Stark et al. data, as a tracer of the Galactic extinction was assessed by Boulanger & Perault (1988) using high resolution far-infrared maps derived from the IRAS satellite. At $100\mu m$, most of the diffuse emission seen by IRAS is assumed to be thermal radiation from dust in the Galaxy. They discovered that at high Galactic latitudes ($b > 50$) the measured HI emission and IRAS 100$\mu$m flux were well correlated, both as a function of Galactic latitude and longitude. Moreover, the correlation was linear thus justifying a constant dust–to–gas ratio. We have used the HI data in preference to the IRAS data because of continuing uncertainties in the zodiacal light subtraction from the IRAS data.

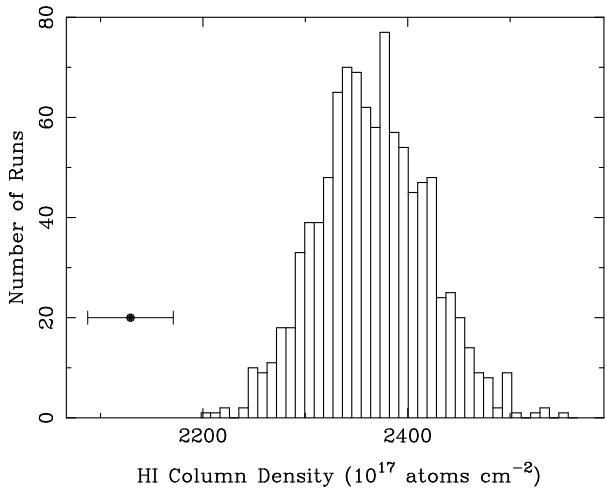

**Figure 2.** Histogram of mean HI column densities for 1000 random runs of the $D > 5$ Abell cluster subsample. Also plotted is the mean determined for the real cluster dataset and the standard error on that measurement. This demonstrates that such a low mean value of HI is not expected from random lines-of-sight.

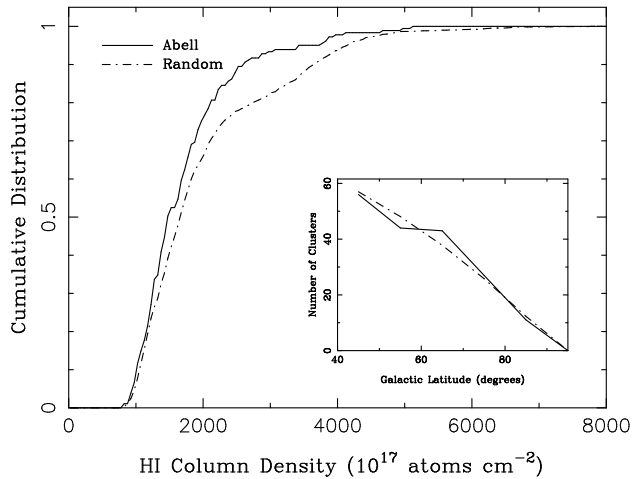

**Figure 3.** The real and random cumulative distributions for the $D > 5$ & $0° \leq l \leq 120°$ Abell subsample. The inset plot again shows the Galactic latitude distributions for the Abell clusters and the random dataset.

For various selections of Abell clusters (see Table 1), the interpolated HI Stark et al. values towards all the clusters were computed and the observed distribution of these HI values was then compared to the observed HI distribution for randomly selected lines-of-sight. In all cases, the number of random directions was twenty times greater than that of the real data and they were selected under the same coordinate constraints as the real data discussed above. In addition, the random directions were constrained to follow the same observed Galactic latitude dependence which was achieved by fitting the observed surface density of clusters for each of the different selections in Table 1 (see Fig. 1). This ensured that the random catalogues covered the same part of the sky and included the same, known selection bias.

The real and random HI cumulative distributions for each of the different cluster subsamples were compared using a Kolmogorov-Smirnov (KS) test and a probability derived that the two distributions were drawn from the same parent distribution. The results of these KS tests are shown in Table 1, while Fig. 1 shows two examples of the real and random cumulative distributions.

As a further test, the mean HI column density and standard error were calculated for each of the samples of Abell clusters given in Table 1. In addition, for each sample, a distribution of random mean HI values was constructed by randomly sampling the HI data under the same coordinate constraints and using the same number of clusters as the real data sample. A demonstration of this is shown in Fig. 2. In all cases, the distribution of random samplings was Gaussian, as expected from the Central Limit Theorem. The significance of the observed Abell HI mean value can then be derived from the mean and standard deviation of these generated distributions.

## 4 DISCUSSION

We have constrained the analysis presented here to areas of the sky believed to be free of extinction problems. However, it is clear from Table 1 that for most of the Abell subsamples considered the distribution of clusters appears to be anti–correlated with respect to the implied distribution of Galactic extinction. All the samples examined here have a lower mean HI column density than that computed for random lines-of-sight. The statistical significance of these departures from random expectations varies between the samples.

Taking all Abell clusters above $b > +40$, irrespective of their richness ($R$) and distance class ($D$; see Abell 1958 for the full definition of these), it appears that their positions are significantly anti-correlated with respect to the implied HI column density. This is not due to the known strong Galactic latitude selection function since this has already been incorporated into the analysis. If we now investigate this anti-correlation as a function of richness (Table 1), the smaller $D \leq 5$ and $R \geq 2$ sample still shows a significant anti–correlation as defined by the lower than expected mean HI value and the high KS probability (99.6% probability that the real and random HI distributions were not drawn from the same parent distribution). The significance of the anti–correlation diminishes as a function of richness, with the $R = 0$ clusters having a mean HI value and KS probability consistent with a random dataset. In contrast, when clusters with $D > 5$ are investigated, irrespective of richness (most are $R > 0$), it is apparent than a large fraction of the observed anti-correlation between the whole Abell sample and the HI column densities comes from this distant cluster sample. This is consistent with Abell's original hypothesis. Similar trends are also seen if we repeat this analysis only for clusters at Galactic latitudes of $b > +55$, re-enforcing the fact that this is not a Galactic latitude effect.

**Table 1.** The results of comparing the positions of Abell clusters with measured HI column densities from the Stark et al. (1992) survey in units of $10^{17}$ atoms cm$^{-2}$. Various Abell subsamples are considered and presented here along with total number of clusters used in the analysis. The KS probability and mean observed Abell HI are also shown. The last column contains the expected mean HI as derived from randomly sampling the data.

| Sample | N | $P_{KS}$ | $\langle HI \rangle$ | Mean $\langle HI \rangle$ |
|---|---|---|---|---|
| $b > +40$ | 1310 | 0.0080 | $2182 \pm 29$ | $2344 \pm 34$ |
| $b > +40$, $D \leq 5$ & $R \geq 2$ | 85 | 0.0048 | $1906 \pm 89$ | $2265 \pm 109$ |
| $b > +40$, $D \leq 5$ & $R \geq 1$ | 448 | 0.0531 | $2163 \pm 48$ | $2249 \pm 54$ |
| $b > +40$, $D \leq 5$ & $R = 0$ | 250 | 0.4980 | $2345 \pm 73$ | $2459 \pm 85$ |
| $b > +40$, $D > 5$ | 612 | 0.0012 | $2129 \pm 42$ | $2365 \pm 54$ |
| $b > +55$ | 762 | 0.0582 | $1821 \pm 23$ | $1988 \pm 44$ |
| $b > +55$, $D \leq 5$ & $R \geq 1$ | 289 | 0.0241 | $1811 \pm 39$ | $1965 \pm 76$ |
| $b > +55$, $D \leq 5$ & $R = 0$ | 128 | 0.3675 | $1863 \pm 73$ | $1896 \pm 95$ |
| $b > +55$, $D > 5$ | 345 | 0.0345 | $1823 \pm 32$ | $1991 \pm 43$ |
| $b > |30|$, Postman et al. sample | 282 | 0.0134 | $2595 \pm 61$ | $2889 \pm 100$ |
| $b > |30|$, Postman et al. sample, $R = 0$ | 155 | 0.0881 | $2873 \pm 108$ | $3064 \pm 123$ |
| $b > |30|$, Postman et al. sample, $R \geq 1$ | 127 | 0.0212 | $2401 \pm 115$ | $2696 \pm 123$ |
| $b > +40$, $D > 5$ & $0° \leq l \leq 120°$ | 181 | 0.0025 | $1746 \pm 58$ | $2022 \pm 81$ |
| $b > +40$, $D > 5$ & $120° < l \leq 240°$ | 297 | 0.0339 | $1926 \pm 43$ | $2101 \pm 44$ |
| $b > +40$, $D > 5$ & $240° < l \leq 360°$ | 134 | 0.301 | $3097 \pm 110$ | $3142 \pm 119$ |

We have also carried out an analysis on the sample of Abell clusters used by Postman et al. (1992), which is the largest sample of Abell clusters with measured redshifts. The sample constitutes all Abell (1958) northern clusters, irrespective of richness, with an $m_{10} \leq 16.5$ (where $m_{10}$ is the magnitude of the tenth cluster galaxy). This sample has been extensively used to study the large-scale clustering properties of clusters. The findings of our analysis are consistent with those mentioned above for the more generic samples of Abell clusters considered here and indicates that the observed angular distribution of this important Abell subsample maybe influenced by Galactic extinction.

As a further test, the data was cut into three separate Galactic longitude segments (see Table 1) to test for any longitudinal dependence on the observed anti-correlation i.e. was any particular region of the sky responsible for all the signal? We carried out such an analysis using the $D > 5$ subsample as this had the most significant anti-correlation. The $240° < l \leq 360°$ region of the data coincides with the area originally highlighted by Abell and contains an anomalously low surface density of clusters, while the $120° < l \leq 240°$ segment covers an area discussed by Bahcall & Soneira (1982) which appears to have an apparent void of Abell clusters.

Once again, all three segments have a lower observed mean HI value than that expected from random directions, yet the only one with a compelling statistical significance is the $0° \leq l \leq 120°$ region. We present the cumulative real and random distributions for this segment in Fig. 3. This is contrary to what may have been expected in light of the above remarks. Although the $240° \leq l \leq 360°$ region has a high mean extinction value and a low surface density of clusters (as originally highlighted by Abell), our analysis indicates that the positions of the 134 clusters in this region are not preferentially located in areas of lower than average extinction; they are effectively randomly scattered with respect to the extinction. The $120° \leq l \leq 240°$ region has a higher surface density of clusters than expected from random which could be used as an indicator of low extinction. However, our analysis tentatively suggests that the positions of these 297 clusters are anti-correlated with the HI column density. We carried out an analysis for the area bounded by $140° \leq l \leq 240°$ and $30° \leq b \leq 60°$ which coincides with the void discussed by Bahcall & Soneira (1982) and found the same level of significance for the observed anti-correlation. This suggests that extinction may play a part in the distribution of clusters in this part of the sky.

This analysis does indicate the severe problems in using fluctuations in the observed surface density of clusters as an indicator of extinction. It is extremely difficult is separate extinction-induced clustering from real large-scale structure. Furthermore, it highlights potential longitudinal gradients in the distribution of Abell clusters. Such an effect is not normally considered in statistical studies of the cluster distribution and could introduce false large-scale signals if the gradients are caused by extinction.

The work presented here implies subtle anti-correlations between the classifications of Abell clusters and the observed distribution of HI column densities. It is relatively straightforward to understand why the more distant clusters appear to be discovered in regions of lower than average inferred extinction, since in these areas the photographic plates are more likely to reach fainter flux limits thus aiding their detection and classification. The anti-correlation between richer Abell systems and HI column densities is less obvious, but is probably a combination of several factors; an increased efficiency in counting and classifying galaxies in the lower relative extinction regions, an inappropriate background subtraction and a larger Abell radius compared to other clusters of similar intrinsic properties in higher extinction regions. As an example, 0.1 magnitudes of relative extinction can boost the richness of an Abell cluster by upto 20% compared to similar clusters nearby on the sky. Extinction–induced differences in the background galaxy correction can account for approximately 5% of this, while the remainder comes from a larger inferred Abell radius which is proportional to $m_{10}$ and would thus be artificially fainter (smaller radius) in the higher extinction region.

Finally, this work does highlight the problems of extinction in constructing and statistically analysing cluster surveys. In recent years, several groups have begun to construct automated samples of clusters from digitised galaxy counts e.g. Lumsden et al. (1992). Such an automated approach will easily allow the effects of extinction, as derived from independent information, to be included into the selection and classification of clusters thus removing the need to correct for it later. This point is emphasized by carrying out an identical correlation analysis on the automated Edinburgh/Durham Cluster Catalogue (EDCC; Lumsden et al. 1992). Although this catalogue was not explicitly corrected for extinction using external indicators, an initial step in its construction involved the subtraction of a sky frame which represented the large–scale fluctuations (scales of $\sim 1.5°$) seen in the original galaxy distribution (Lumsden et al. 1992). This process certainly helped in reducing the effects of extinction, since the typical coherence length of the patchy extinction is $\simeq 5° \rightarrow 10°$ (Nichol & Collins 1994). No correlation, significant or not, was found between the observed HI column densities and the richnesses, distances or positions of the clusters.

## ACKNOWLEDGMENTS

Thanks to Rich Kron for his initial insight. AJC acknowledges partial support from NSF grant AST-9020380, while we both acknowledge partial support from ADP grant NAG 5-2399. This research has made use of some data obtained through the High Energy Astrophysics Science Archive Research Center Online Service, provided by the NASA-Goddard Space Flight Center. Finally, we are very grateful to Stuart Lumsden, the referee, for his helpful comments on the paper.